\documentclass[acmsmall]{acmart}

\AtBeginDocument{%
  }

\setcopyright{acmlicensed}
\copyrightyear{2024}
\acmYear{2024}
\acmDOI{XXXXXXX.XXXXXXX}

\acmJournal{JACM}
\acmVolume{37}
\acmNumber{4}
\acmArticle{111}
\acmMonth{8}

\usepackage{algorithm}
\usepackage{algpseudocode}
\usepackage{amsmath}
\usepackage{multirow}
\usepackage{colortbl}
\usepackage{pifont}

\begin{document}

\title{SparseMap: Loop Mapping for Sparse CNNs on Streaming Coarse-grained Reconfigurable Array}


\author{Xiaobing Ni}
\email{nxb@mail.ustc.edu.cn}
\affiliation{%
	\institution{School of Microelectronics, University of Science and Technology of China}
	\city{Hefei}
	\state{Anhui}
	\country{China}
}

\author{Mengke Ge}
\email{mengke.ge@iai.ustc.edu.cn}
\affiliation{%
	\institution{Institute of Artificial Intelligence, Hefei Comprehensive National Science Center}
	\city{Hefei}
	\state{Anhui}
	\country{China}
}
\affiliation{%
	\institution{School of Microelectronics, University of Science and Technology of China}
	\city{Hefei}
	\state{Anhui}
	\country{China}
}

\author{Jiaheng Ruan}
\email{handso@mail.ustc.edu.cn}
\affiliation{%
	\institution{School of Microelectronics, University of Science and Technology of China}
	\city{Hefei}
	\state{Anhui}
	\country{China}
}

\author{Song Chen, Yi Kang}
\email{songch, ykang@ustc.edu.cn}
\affiliation{%
	\institution{School of Microelectronics, University of Science and Technology of China}
	\city{Hefei}
	\state{Anhui}
	\country{China}
}
\affiliation{%
	\institution{Institute of Artificial Intelligence, Hefei Comprehensive National Science Center}
	\city{Hefei}
	\state{Anhui}
	\country{China}
}

\thanks{ This work was supported in part by the Strategic Priority Research
	Program of Chinese Academy of Sciences, Grant No. XDB44000000,
	in part by the University Synergy Innovation Program of Anhui
	Province under grant No. GXXT-2023-003, and in part by CAS
	Project for Young Scientists in Basic Research under grant No. YSBR-029.  \emph{(Corresponding author: Song Chen)}
}

\renewcommand{\shortauthors}{Ni et al.}

\begin{abstract}
Streaming coarse-grained reconfigurable array (CGRA)  is a promising architecture for data/computing-intensive applications because of its flexibility, high throughput and efficient memory system. However, when accelerating sparse CNNs, the irregular   input data demands inside sparse CNNs would  cause excessive caching operations (COPs) and  multi-cycle  internal dependencies (MCIDs) between operations, declining the throughput of the streaming CGRA. We propose a mapping method for sparse CNNs onto streaming CGRA, SparseMap, which incorporates an efficient I/O data management along  with operation scheduling  and binding, to reduce the COPs and MCIDs, thereby ensuring the optimal throughput of streaming CGRA.
The experimental results show SparseMap  reduces 92.5\% COPs and 46.0 \% MCIDs while achieves the same or even smaller initiation interval (II) compared to previous works.
\end{abstract}

\ccsdesc[500]{Computer systems organization~Reconfigurable computing}
\ccsdesc[300]{Computing methodologies~Parallel computing methodologies}
\ccsdesc[100]{Theory of computation~Models of computation}

\keywords{Streaming CGRA,  sparse CNNs, irregular input demands,  COPs, MCIDs,   SparseMap, I/O data management.}

\received{20 February 2007}
\received[revised]{12 March 2009}
\received[accepted]{5 June 2009}

\maketitle

\section{Introduction}
CGRA, combining energy efficiency and flexibility, is a promising architecture for data/computing-intensive applications\cite{softbrain,7983395, MemOptimization, Arch}. 
Inspired by  the streaming  access property inherent to  data/computing-intensive applications, a streaming CGRA was proposed \cite{softbrain}, where  on-chip data memories  provide a streaming dataflow  for the PE array (PEA), avoiding the fragmented and scattered memory access from PEs. 
In comparison to the common CGRAs utilizing load/store addressing in PEs, the streaming CGRA is more conducive to applications that employ continuous memory addressing, such as CNNs.  The streaming CGRA in Fig. 1 is equipped with a  PEA,  a global register file (GRF), data memories ( supporting streaming access), a crossbar and I/O buses. 

\begin{figure}[t]
	\centering
	\includegraphics[width=0.9\linewidth]{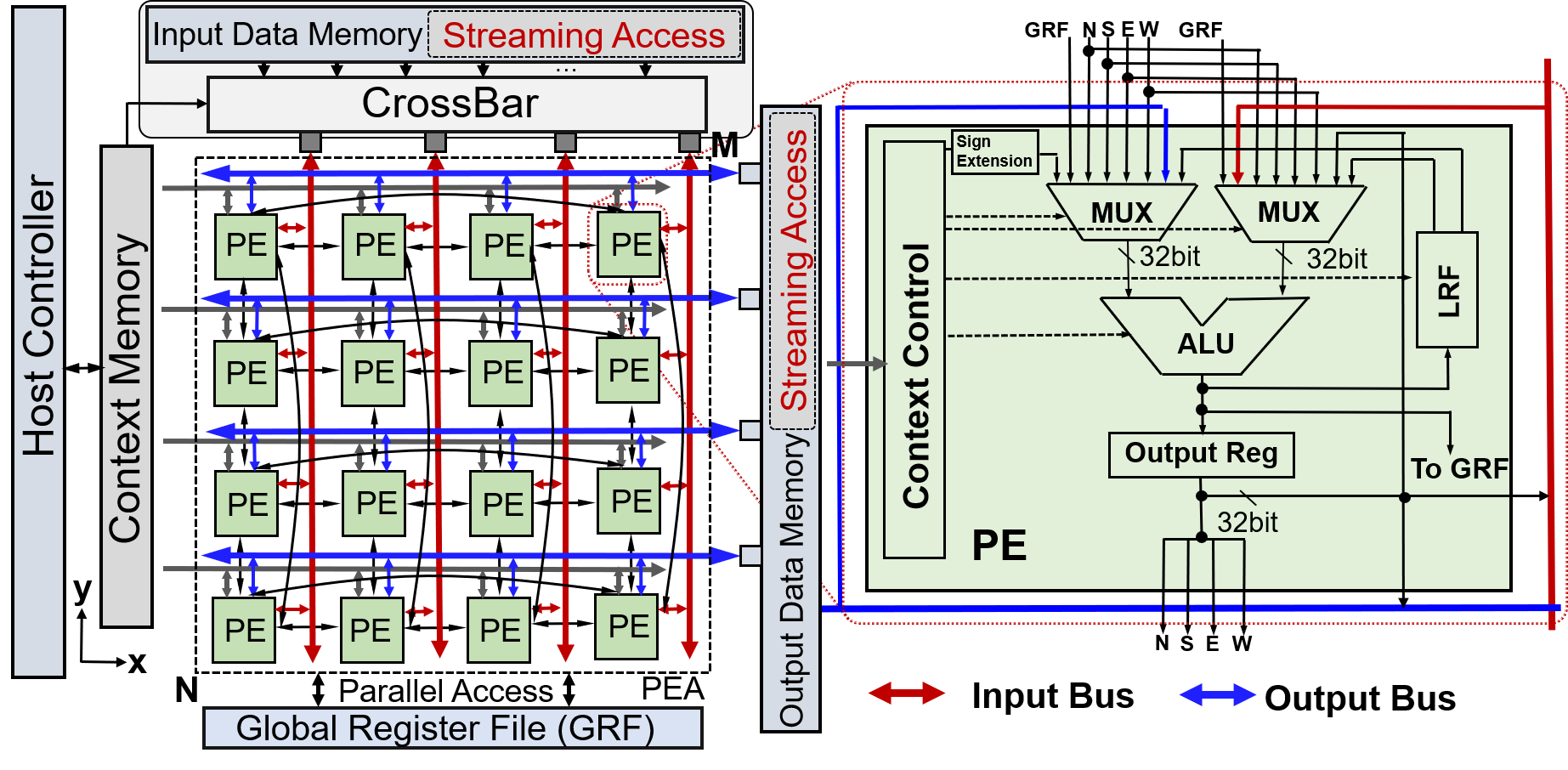}
	\Description{A streaming CGRA architecture along with the PE microarchitecture}
	\caption{Streaming CGRA.}
\end{figure}

The CGRA accelerates the applications' kernel loops. The  CGRA compiler performs soft pipelining for the loop,  schedules  loop operations, and binds the operations and internal dependencies  to PEs and routing  resources\cite{hamzeh_regimap_2013}.
Previous compilers can be categorized into graph based\cite{BusMap,kou_taem_2020,hamzeh_regimap_2013,kou2022geml, yin_memory-aware_2016,dave2018ramp,hamzeh_epimap_2012}, cost function based \cite{zhao_towards_2020,friedman2009spr,wijerathne2022panorama, EdgeCentric} and learning based\cite{MapZero}\cite{E2E} methods. These works aimed to minimize the II between adjacent iterations of the loop to maximize the throughput of CGRA. They focused on  accelerating the applications on  common CGRAs. However, due to the absence of  load/store addressing in PEs and no buffer in I/O buses, the compiler for streaming CGRA  needs to explicitly manage the I/O data at the PEA side, i.e.  allocating I/O data with  the correct I/O buses at the appropriate time  and caching the I/O data in the PEA if needed,  thus ensuring the legal I/O data routing.

CNNs outperform humans \cite{VGG}\cite{AlexNet} but with  surging computation and memory requirements\cite{AI}. 
Fortunately, the intrinsic sparsity in CNNs  significantly lowers these  requirements in sparse CNNs\cite{DeepComp3}\cite{DeepComp4}.  The sparse CNN is typically partitioned into multiple sparse blocks which are handled in a predetermined order.  Each block computes different channels from different kernels and  is abstracted as a sparse data flow graph (\emph{\textbf{s-DFG}}).  
The non-uniform distribution of non-zero weights  and the skipping of zero-weight multiplications would  incur irregular  input data demands inside s-DFGs. The irregularity complicates the management of input data  at the PEA side,  and further affects the efficiency of  the streaming CGRA. Firstly, due to no buffer in I/O buses,  PEs would be occupied to cache  the input data  whose allocating time is inconsistent with the scheduling times  of its multiplications. Secondly,  without proper  bus allocation for input data, the excessive MCIDs  between operations would be generated.  Limited by the PEA and buffer resources, the excessive PEs for caching  (corresponding to the COPs inserted into s-DFG)  and  MCIDs routing may increase  the II, thereby declining   streaming CGRA's throughput.  

\begin{figure}[t]
	\centering
	\includegraphics[width=0.72\linewidth]{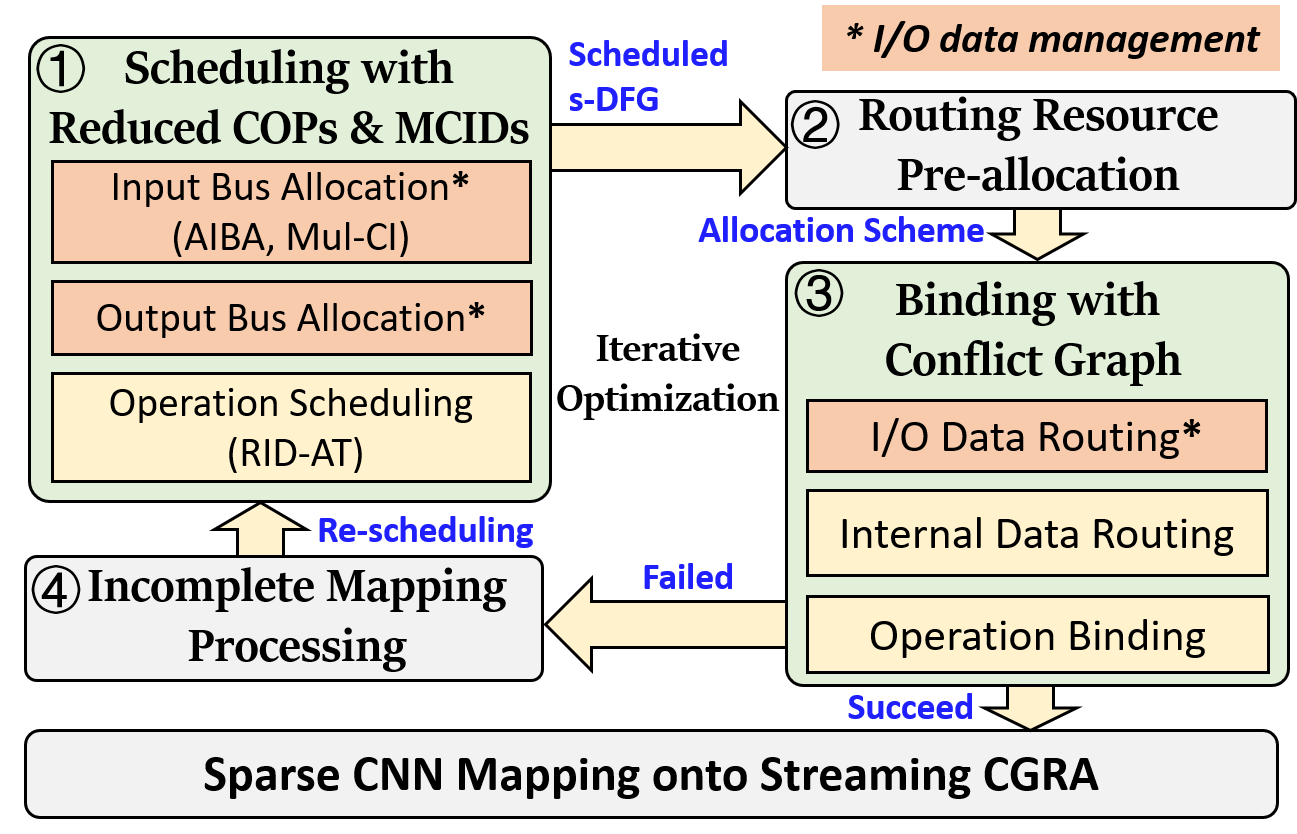}
	\Description{SparseMap overview contains four phases: scheduling with reduced COPs and MCIDs, routing resources pre-allocation, binding with conflict graph and incomplete mapping processing.}
	\caption{SparseMap overview.}
\end{figure}

In order to address the issue of excessive COPs and MCIDs  resulting from the irregular input data demands, we propose SparseMap, a mapping algorithm designed for sparse CNNs in the context of streaming CGRAs. SparseMap employs an efficient I/O data management, along with operation scheduling and binding, to ensure the optimal throughput of streaming CGRAs. It contains contributions as follows:

\begin{itemize}
	\item To address the challenges posed by the irregular input data demands inside the s-DFG, we present a scheduling approach that not only allocates input buses in accordance with input data demands but also reconstructs the internal dependencies within the adder trees. This enables the s-DFG to be scheduled in a manner that reduces both COPs and MCIDs.
	\item  Given the scheduled s-DFG and streaming CGRA, we propose to construct a conflict graph to represent the binding problem, such that legal operation binding, and  I/O data and internal data routing are obtained by solving the maximum independent set (\emph{\textbf{MIS}}) on the graph.
	\item Compared to previous works, SparseMap achieves fewer COPs and MCIDs while having the same or even smaller II.
\end{itemize}

The rest of this paper is organized as follows: Section 2 elaborates the way to reduce COPs and MCIDs. Section 3 gives the problem formulation. Section 4 details  SparseMap. Section 5 presents experimental results, followed by the conclusion in Section 6.

\section{Reducing COPs and MCIDs}
To reduce the COPs and MCIDs caused by the irregular input data  demands, we allocate the input buses   considering the demands  among multiple input data and within single input data.  Moreover,  the internal dependencies within the adder trees are to be reconstructed to further reduce the MCIDs.

\subsection{Association-Oriented Input Bus Allocation}
Any two input data with a high \textbf{association} ( defined as the number of kernels requiring these two data simultaneously) allocated with input buses at  different  times  would incur excessive MCIDs. Firstly, to avoid the caching PEs, for any input data, the scheduling times of its multiplications should keep same as the allocating time of that data. Secondly, for any addition, at least 1 MCID is generated if its two producers (multiplications or additions) are scheduled at different times. Hence,  any two  highly associated input data with different allocating times would incur their respective multiplications   to be scheduled differently, thereby causing the excessive MCIDs.  

We propose an \underline{a}ssociation-oriented   \underline{i}nput \underline{b}us \underline{a}llocation (\emph{\textbf{AIBA}}) method which prioritizes  allocating input buses for highly associated input data at the same time, making more  multiplications be scheduled together across different kernels, thus reducing the MCIDs. 
The efficiency of AIBA can be proven in Fig. 3 where the input data $c_2$ and $c_3$ with a highest association are allocated with input buses at the same time in Fig. 3(d), reducing the MCIDs from 3 in Fig. 3(c) to 1. Routing  via  local register file (LRF) for any MCID in Fig. 3(c)-(d) is forbidden due to the same modulo time for the consumer and producer in each MCID\cite{BusMap}. Routing via GRF can be adopted, but is able for  1 MCID at most, causing a failed mapping at  II = 2 for  Fig. 3(c), but a successful mapping for  Fig. 3(d). The reduced MCIDs alleviate the routing pressures and ensure the streaming CGRA's throughput.

\begin{figure}[t]
	\centering
	\includegraphics[width=0.9\linewidth]{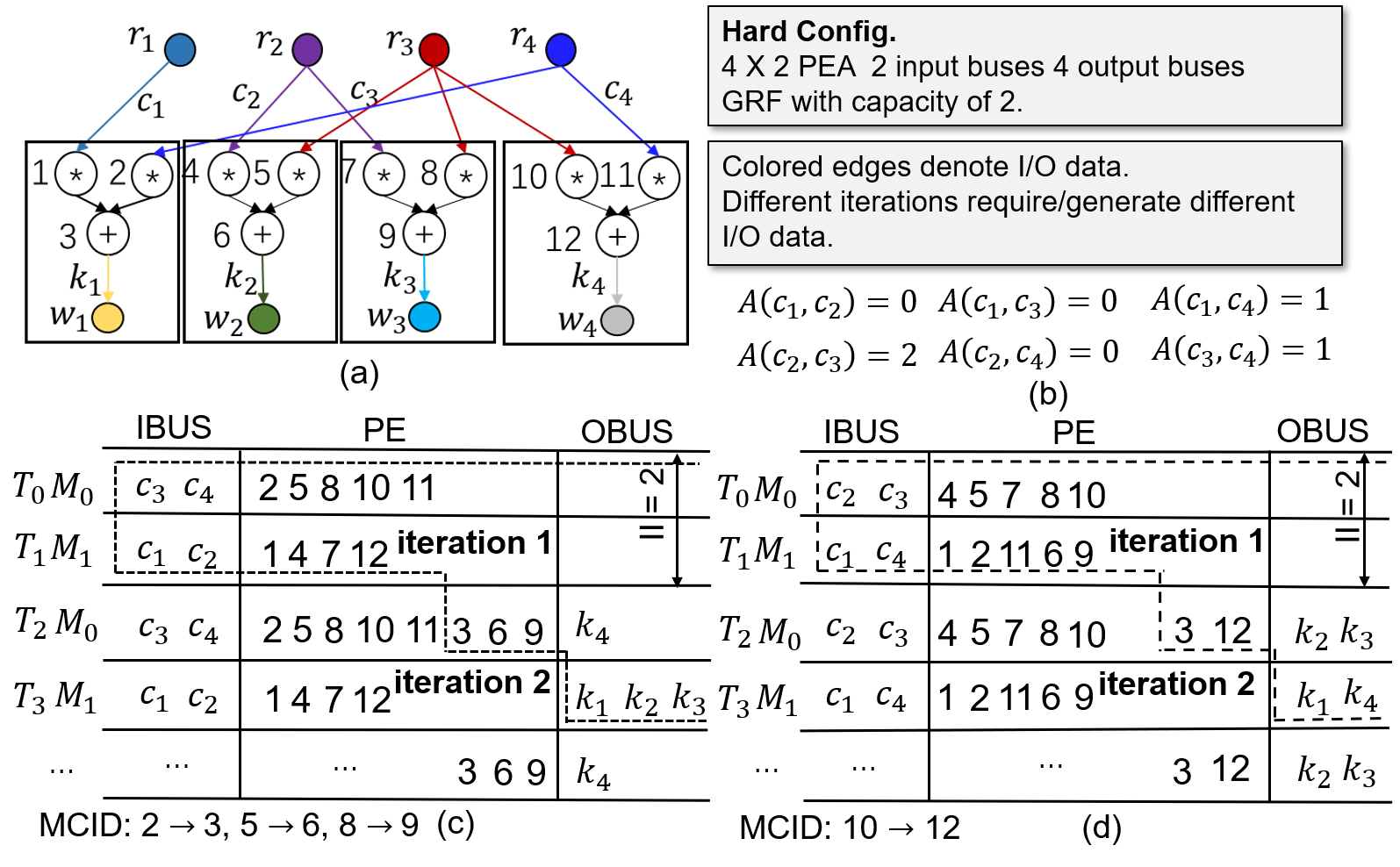}
	\Description{For a s-DFG computing 4 channels from 4 kernels, different input bus allocations lead to different MICDs.}
	\caption{(a) A s-DFG computing 4 channels from 4 kernels; (b) Input associations; (c) A scheduling  with 3 MCIDs; (d) A scheduling  with 1 MCID.}
\end{figure}

\begin{figure}[t]
	\centering
	\includegraphics[width=0.45\linewidth]{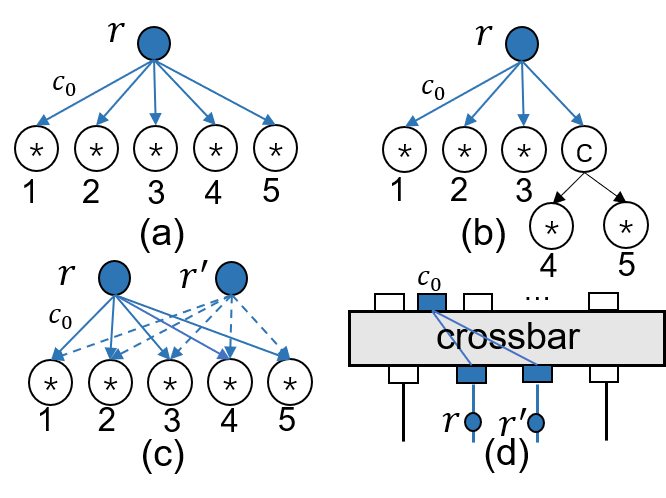}
	\Description{Multi-casting the input data with high fanout multiplications can avoid the caching operations.}
	\caption{(a) The input data $c_0$ with 5 multiplications exceeding fan-out PEs on an input bus in a 4$\times$4 PEA; (b) A COP $c$ inserted into s-DFG; (c) (d) Multi-casting $c_0$  to 2 input buses by crossbar.}
\end{figure}

\subsection{Multi-casting  Input Data via Crossbar}
For the input data with multiplications exceeding the fan-out PEs of an input bus, not all the PEs executing the multiplications can   obtain the input data from one input bus directly (the weights are pre-loaded into PEs' LRF). In this case,   not all the  multiplications can be scheduled at the allocating time of the input data. As Fig. 4(a)-(b) show, on a 4 $\times$ 4 PE, the input data $c_0$ cannot be directly transferred to 5 PEs executing its multiplications via one input bus, and a COP is required for $c_0$.  Aided by the multi-casting ability of the crossbar,   we propose to \underline{mul}ti-\underline{c}ast the \underline{i}nput data (\emph{\textbf{Mul-CI}}) to more input buses (if available) as depicted in Fig. 4(c)-(d), which avoids the COPs and enables more multiplications to be scheduled with the input data. The Mul-CI technique also indirectly guarantees the effectiveness of the AIBA technique.

\begin{figure}[t]
	\centering
	\includegraphics[width=0.75\linewidth]{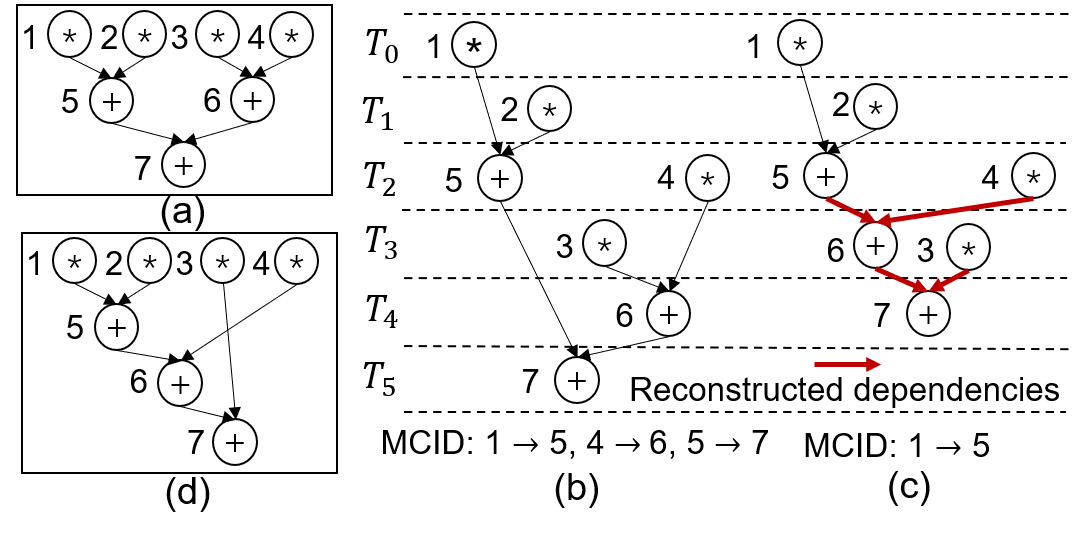}
	\Description{Reconstructing the internal dependencies with the adder tree according to the scheduling of the multiplications can reduce the MCIDs.}
	\caption{(a) A kernel with 4 multiplications and 1 adder tree composed by 3 additions; (b) A scheduling with 3 MCIDs for the fixed adder tree. (c) A scheduling with 1 MCID for the reconstructed adder tree; (d) The final kernel after reconstructing the internal dependencies within the adder tree.}
\end{figure}

\subsection{Reconstructing Internal Dependencies within Adder Trees}
For a kernel with $n$ multiplications, $n$ - 1 additions form an adder tree to accumulate the multiplications. As long as it is guaranteed that there are  two producers and one consumer for each addition (except for the last addition), the final accumulated result is unique and correct regardless of the internal dependencies within the adder tree. Based on the observation above, we  propose to \underline{r}econstruct the  \underline{i}nternal \underline{d}ependencies within the \underline{a}dder \underline{t}ree (\emph{\textbf{RID-AT}}) according to the multiplications' scheduling to further reduce the MCIDs. Section \uppercase\expandafter{\romannumeral 4} details the RID-AT technique.  
After employing RID-AT for the kernel in Fig. 5(a), the  MCIDs in Fig. 5(c) are reduced from 3 to 1 compared with the fixed adder tree in Fig. 5(b).

\section{Problem Formulation}

\subsection{Definitions and Notations}
The mapping of s-DFG onto streaming CGRA can be decomposed into scheduling and binding\cite{hamzeh_regimap_2013}\cite{BusMap}. 
Targeting the reduction of COPs and MCIDs, the scheduling determines the scheduling and modulo scheduling times of operations and the I/O bus allocation for the I/O data, subject to the constraints dependencies, PEA and I/O buses.  The binding decides  the physical resource required for each node in s-DFG and ensures the legal routing of I/O data and internal data.
To clearly define the s-DFG mapping onto  streaming CGRA,   we introduce I/O reading/writing nodes into s-DFG to facilitate the I/O data  management, along with operation scheduling and binding.

(1) \emph{Sparse Data Flow Graph (s-DFG):} s-DFG is the  loop body of the sparse block, denoted as $D$ = $(V_D, E_D)$. $V_D$ is  a  set composed of multiplications, additions and I/O  reading/writing which constitute $V_{M}$, $V_A$, $V_{R}$ and $V_{W}$ respectively, and we denote an operation set $V_{OP}$ = $V_M$ $\cup$ $V_A$. Differing from the nodes in $V_{OP}$ executed by PEs, the nodes in $V_R$ ($V_W$) are operated on input  (output) buses, denoting the input (output) data is read from (written to) input (output) buses.  $E_D$ is a dependency set containing input dependencies, output dependencies, and internal dependencies, denoted as $E_D$ = $E_R$ $\cup$ $E_W$ $\cup$ $E_I$: $\forall$ ($v_1$, $v_2$) $\in$ $E_D$, if $v_1$ $\in$ $V_R$ , $v_2$ $\in$ $V_{OP}$, then ($v_1$, $v_2$) $\in$ $E_R$; if $v_1$ $\in$ $V_{OP}$ , $v_2$ $\in$ $V_{W}$, then ($v_1$, $v_2$) $\in$ $E_W$; otherwise, ($v_1$, $v_2$) $\in$ $E_I$. 
The internal dependencies within the adder trees would  be reconstructed in the scheduling phase.

(2) \emph{Initiation Interval (II):} II is the time slots between adjacent iterations of the loop. 
The compiler pursues to minimize II to maximize the throughput of the  streaming CGRA. 
After scheduling, $\forall$ $v$ $\in$ $V_D$ is assigned a scheduling time and a modulo scheduling time  from 0 to II -1. The (modulo) scheduling time  of  $r$ ($w$) $\in$ $V_R$ ($V_W$)  denotes the (modulo) allocating time of the I/O data.

(3) \emph{I/O data management:} The management of I/O data at PEA side can be transformed into the scheduling, caching and binding  for the nodes in $V_R$ and $V_W$. Due to no buffer in I/O buses, the scheduling distance between the consumer and producer of any input (output) dependency can only be 0 (1).

(4) \emph{Time-Extended CGRA (TEC):} TEC $T$ = $(V_T, E_T, II)$ is a resource graph replicating streaming CGRA from 0 to II-1. $V_T$ is a  set composed of replicated resource nodes (including PEs, I/O buses etc.). $\forall$ $v^m$ $\in$ $V_T$ denotes the resource node $v$ located at $m$-th time layer of the TEC.  $E_T$ is a directed edge set. For  $v_1^{m_1}$, $v_2^{m_2}$ $\in$ $V_T$, $v_1^{m_1}$   connects to $v_2^{m_2}$ if and only if $(v_1^{m_1}$, $v_2^{m_2})$ $\in E_T$, $m_2$ = $m_1$ + 1, $0\leq m_1 <$ II - 1, or $m_2$ = 0, $m_1$ = II -1\cite{hamzeh_regimap_2013}\cite{BusMap}.

(5) \emph{Conflict Graph (CG) and MIS:} CG $CG = (V_{CG}, E_{CG})$ is a graph  reflecting  the conflict relations among  bindings. $V_{CG}$ contains: tuples $(r^m, ibus_i^m)$, $(w^m, obus_j^m)$ and quadruples $(pe^{m}_{i,j}, op^m, bus^m_{x}, bus^m_{y})$.  $(r^m, ibus_i^m)$  denotes the input reading $r$ $\in$ $V_R$ modulo scheduled at $m$ is allocated with $i$-th input bus at the $m$-th time layer of the TEC, so as $(w^m, obus_j^m)$ dose.  $(pe^{m}_{i,j}, op^m, bus^m_{x}, bus^m_{y})$ denotes $op$ $\in$ $V_{OP}$ modulo scheduled at  $m$ is executed by the PE instance located at  position $(i,j)$ at the $m$-th time layer. 
$bus^m_{x}$ and $bus^m_{y}$ determine the bus routing among PEs \cite{BusMap}. Each edge in $E_{CG}$ denotes a resource conflict between the two bindings which cannot coexist in any independent set of CG. Section 4  details the construction of  CG. The MIS of CG means the most bindings without conflict.

\begin{table}[t]
	\centering
	\captionsetup{font=footnotesize}\caption{Symbols and descriptions}
	{
		\begin{tabular}{ll}
			\toprule
			$N$, $M$         &  $N$ $\times$ $M$ PEA with $M$ input buses, $N$ output buses. \\     
			$t(v)$, $m(v)$   &  The scheduling and modulo scheduling times for $v$ $\in$ $V_D$. \\
			$c_r$            &   0-1 variable. $c_r$ is 1 if  $r$ $\in$ $V_R$ is cached by a COP. \\
			$mc_r$           &   0-1 variable. $mc_r$ is 1 if   $r$ $\in$ $V_R$  is multi-cast by crossbar.\\
			$MCID$           &  \{ ($v_1$, $v_2$) $|$ ($v_1$, $v_2$) $\in$ $E_I$,  $t(v_1)$ - $t(v_2)$ $>$ 1 \}, denoting the set of MCIDs.\\  
			$\mathcal{M}_R(i)$       &  \{ $r$ $|$ $r$ $\in$ $V_R$, $m(r)$ = $i$ \}, composed of the $r$ $\in$ $V_R$  modulo scheduled at $i$.\\
			$\mathcal{M}_W(i)$       &  \{ $w$ $|$ $w$ $\in$ $V_W$, $m(w)$ = $i$ \}, composed of the $w$ $\in$ $V_W$ modulo scheduled at $i$.\\
			                         
		    $\mathcal{M}_{OP}(i)$    &  \{ $op$ $|$ $op$ $\in$ $V_{OP}$, $m(op)$ = $i$ \}, composed of the $op$ $\in$ $V_{OP}$ modulo scheduled at $i$. \\
			$\mathcal{M}_C(i)$       & \{ $r$ $|$ $c_r$ = 1, $m(r)$ = $i$ \}, composed of  the $r$ $\in$ $V_R$ which  is   cached by a COP.    \\
	
			$\mathcal{M}_{MC}(i)$    & \{ $r$ $|$ $mc_r$ = 1, $m(r)$ = $i$ \}, composed of  the $r$ $\in$ $V_R$ which  is  multi-cast by crossbar.\\
			\bottomrule
		\end{tabular}
	}
\end{table}

\subsection{Problem Definition}
The scheduling determines the scheduling and modulo scheduling  times for each node in $V_D$, where the modulo scheduling time of each node indicates the time layer of the TEC when that node is operated.  In Table 1, we list all the symbols used in the following texts. The scheduling  can be defined as finding  a scheduling \{ $m(v)$ $|$ $m(v)$ = $t(v)$ \% II,  $v$ $\in$ $V_D$\},  such that: 
	\begin{itemize}
		\item[(1)] $\forall$ $(v_1, v_2)$ $\in$ $E_R$, $t(v_2)$ = $t(v_1)$;   $\forall$ $(v_1, v_2)$ $\in$ $E_W$, $t(v_2)$ = $t(v_1)$ + 1; $\forall$ $(v_1, v_2)$ $\in$ $E_I$,  $t(v_2)$ - $t(v_1)$$\geq$1;
		\item[(2)] $\forall$ $i$ from 0 to II - 1, $|\mathcal{M}_R(i)|$ + $|\mathcal{M}_{MC}(i)|$ $\leq$ $M$; $|\mathcal{M}_W(i)|$ $\leq$ $N$;  $|\mathcal{M}_{OP}(i)|$ + 	$|\mathcal{M}_C(i)|$ $\leq$ $N$ $\times$ M; 
		\item[(3)] Minimizing II;   
		\item[(4)] On the premise of (3), minimizing $\sum_{i = 0}^{II -1}$$|\mathcal{M}_C(i)|$ and  $|MCID|$.
	\end{itemize}

(1) and (2) respect the constraints of dependencies and modulo resources. (3) is the top concern, i.e. minimizing II to maximize the streaming CGRA throughput. On the promise of minimizing II, we pursue to minimize the COPs and MCIDs to ensure the optimal throughput of  streaming CGRA.

 The binding decides the physical resources that operate the nodes in $V_D$ at the  corresponding time layers of the TEC without resource conflicts. The binding can be defined as finding a binding set $B$ = \{$b_1$, $b_2$, ..., $b_k$\} $\subset$ $V_{CG}$ of the conflict graph $CG$, such that:  
\begin{itemize}
	\item [(1)] $\forall$ $b_i$, $b_j$ $\in$ $B$, then ($b_i$, $b_j$) $\notin$ $E_{CG}$, ($b_j$, $b_i$) $\notin$ $E_{CG}$;
	\item [(2)] $|B|$ = $|V_D|$
\end{itemize}
 
 Since any edge in $E_{CG}$ denotes a resource conflict between its two bindings, (1) and (2) guarantee a valid mapping, i.e, all the nodes in $V_D$ are bound with the resources on the streaming CGRA without resource conflict.
 
\section{SparseMap Algorithm}
We propose SparseMap, a mapping algorithm  for sparse CNNs onto streaming CGRA, which incorporates an efficient I/O data management  with  operations scheduling and binding,  thereby ensuring optimal throughput for streaming CGRA. 
 
SparseMap  contains four phases shown in Fig. 2.
\ding{172} The scheduling phase achieves the s-DFG's  scheduling with reduced COPs and MCIDs by applying AIBA, Mul-CI and RID-AT techniques.
 To reduce the mapping passes, \ding{173} we  pre-allocate routing resources for the internal dependencies.  \ding{174} We then complete the operations binding and the I/O data and internal data routing by solving the MIS on  a conflict graph. Finally, \ding{175} SparseMap handles the incomplete mapping   if it occurs. 
 
 Our previous work BusMap\cite{BusMap} utilized the  buses to route the internal data among PEs, and achieved reduced GRF access and inserted operations. BusMap focused on accelerating the applications on  common CGRAs, but neglected the impact of I/O data management on the throughput of   CGRA.  In this paper, we implement an efficient I/O data management  in \ding{172} and \ding{174}, and applies the mechanisms described in BusMap  for phases \ding{173} and \ding{175}.

 \begin{algorithm}[t]
 	\caption{Scheduling with Reduced COPs \& MCIDs}\label{alg:alg1}
 	\hspace{-7.0cm}\textbf{Input:} s-DFG D($V_D$, $E_D$), Streaming CGRA C\\
 	\hspace{-7.7cm}\textbf{Output:} Scheduled s-DFG D$'$($V_D'$, $E_D'$), II
 	\begin{algorithmic}[1]
 		\State MII $\gets$ CalculateMII(D, C); II $\gets$ MII;  Scheduling time t $\gets$ 0;  
 		\State Modulo resource tables T$_{PE}$[] $\gets$ 0; T$_I$[] $\gets$ 0; T$_O$[] $\gets$ 0; 
 		\State Unscheduled input readings U$_R$ $\gets$ V$_R$; 
 		\State \textbf{while}  U$_R$ $\neq$ $\emptyset$ \textbf{do}
 		\State \hspace{0.2cm}  Modulo scheduling time m $\gets$ t \% II;
 		\State \hspace{0.2cm} \textbf{if} T$_I$[m] + 1 $>$ $N$ \textbf{then} 
 		\State \hspace{0.2cm} \hspace{0.2cm} t $\gets$ t + 1; 
 		\State \hspace{0.2cm} \hspace{0.2cm} \textbf{continue};
 		\State \hspace{0.2cm} \textbf{end if}
 		\State \hspace{0.2cm} $r$ $\gets$ AIBA(V$_R$, U$_R$, t); T$_I$[m]++; $r$.assign(t, m); U$_R$ $\gets$ U$_R$ - \{$r$\};
 		\State \hspace{0.2cm} \textbf{if} $|$fanout($r$)$|$ + T$_{PE}$[m] $\leq$ $N \times M$ \textbf{then}	
 		\State \hspace{0.2cm}  \hspace{0.2cm} \textbf{if} $|$fanout($r$)$|$ $\leq$ $N$  \textbf{then}
 		\State \hspace{0.2cm}  \hspace{0.2cm} \hspace{0.2cm}  T$_{PE}$[m] $\gets$ T$_{PE}$[m] + $|$fanout($r$)$|$; 
 		\State \hspace{0.2cm}  \hspace{0.2cm} \hspace{0.2cm}  fanout($r$).assign(t, m); 
 		\State \hspace{0.2cm}  \hspace{0.2cm} \hspace{0.2cm}  \textbf{continue};
 		\State \hspace{0.2cm}  \hspace{0.2cm} \textbf{end if} 
 		\State \hspace{0.2cm}  \hspace{0.2cm} \textbf{if} Mul-CI(fanout($r$), T$_{PE}$, T$_I$, t, m) = \textbf{true}  \textbf{then} 
 		\State \hspace{0.2cm}  \hspace{0.2cm} \hspace{0.2cm} \textbf{continue}; 
 		\State \hspace{0.2cm}  \hspace{0.2cm} \textbf{end if} 
 		\State \hspace{0.2cm}  \hspace{0.2cm} \textbf{if}  SchedwithCaching(fanout($r$), T$_{PE}$, t, m) = \textbf{true} \textbf{then}  
 		\State \hspace{0.2cm}  \hspace{0.2cm} \hspace{0.2cm} \textbf{continue};
 		\State \hspace{0.2cm}  \hspace{0.2cm} \textbf{end if} 
 		\State \hspace{0.2cm}  \hspace{0.2cm}  II $\gets$ II + 1; \textbf{goto 2};
 		\State \hspace{0.2cm} \textbf{else if} SchedwithCaching(fanout($r$),  T$_{PE}$, t, m) = \textbf{true} \textbf{then} 
 		\State \hspace{0.2cm}  \hspace{0.2cm} \textbf{continue};
 		\State \hspace{0.2cm} \textbf{end if}
 		\State \hspace{0.2cm} II $\gets$ II + 1; \textbf{goto 2};
 		\State \textbf{end while}
 		\State SchedRemainMulti();
 		\State RID-AT(D, C, T$_{PE}$, II); 
 		\State $D'$ $\gets$ SchedWriting(D, C, T$_{PE}$, T$_O$, II);
 		\State \textbf{return} $D'$, II;

 	\end{algorithmic}
 \end{algorithm}

\subsection{Scheduling with Reduced COPs and MCIDs}
 \ding{182} \textbf{Algorithm overview.}  We propose a scheduling approach in Algorithm 1 that consists of three main procedures.  (1) To reduce the COPs and MCIDs, lines 4-28 complete the scheduling of  input readings and its fanout multiplications while employing the AIBA and Mul-CI techniques. (2) Line 30 applies the RID-AT technique for each kernel to further reduce the MCIDs. (3) Line 31 finishes the scheduling of the output writings.

To minimize II,  we start from the minimum II (MII) =  $max$($\lceil\frac{|V_{OP}|}{N \times M}\rceil$, $\lceil\frac{|V_R|}{M}\rceil$, $\lceil\frac{|V_W|}{N}\rceil$) as  lines 1 shows. Lines 4-28 iteratively allocate input buses for each input reading $r$ $\in$ $V_R$ and schedule its fan-out multiplications $fanout(r)$.  In line 10, the input reading $r$ highly \emph{associated} with scheduled input reading set  is chosen and allocated with an input bus.   For the $r$ with fanouts exceeding the available modulo PEs,  $fanout(r)$ would be scheduled with a caching operation in line 24.
If the modulo PEs are sufficient and  $|fanout(r)|$ is less than the amount of fan-out PEs on an input bus,  $fanout(r)$ would be scheduled at $t$ directly. Otherwise,  in line 17, we  allocate more input buses for $r$. Concretely, we insert a new input reading node $r'$ into the s-DFG, and establish the input dependencies  between $r'$ and the fanout multiplications of $r$ as Fig. 4(c) show. $r'$ and  $fanout(r)$ would be schedule at $t$. The creation of $r'$ and the establishment of input dependencies above model the multi-casting of $r$. If the Mul-CI loses efficacy, we  try to schedule $fanout(r)$ with a caching operation (line 20).  The modulo resource tables  $T_{PE}$ and $T_{I}$ are updated along with the scheduling above. The s-DFG would be rescheduled under increased II if the scheduling for $fanout(r)$ is failed.  After all input readings being scheduled,  we then schedule the remained multiplications in line 29. Finally,  we reconstruct internal dependencies for the adder tree in each kernel and schedule   each output writing in lines 30-31.

\ding{183} \textbf{RID-AT.}  To further reduce the MCIDs,   the internal dependencies within  the adder tree for each kernel are greedily reconstructed according to the multiplications' scheduling. Assume that $t_0$ is earliest scheduling time among  the multiplications.  As soon as two  unaccumulated operations (multiplications or additions)  have  been scheduled before $t_1$ = $t_0$ + 1 and modulo PEs are available at $t_1$, we schedule an addition $v_a$ at $t_1$, and then construct the internal dependencies between these two operations and  $v_a$. Then these two operations and $v_a$ are set as accumulated and unaccumulated  statuses respectively. Otherwise, $t_0$ = $t_0$ + 1, repeat the process above until only one unaccumulated addition is left.  Fig. 6 presents  the snapshots of RID-AT  for the kernel in Fig. 5.

\begin{figure}[t]
	\centering
	\includegraphics[width=0.75\linewidth]{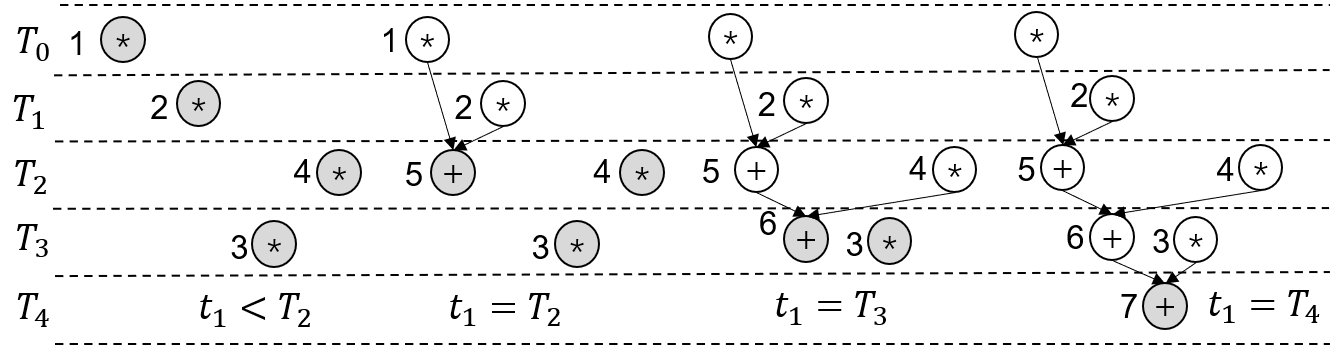}
	\Description{Four snapshots when reconstructing the internal dependencies with the adder tree in Fig. 5.}
	\caption{Snapshots of RID-AT. Grey nodes are unaccumulated.}
\end{figure}

\ding{184} \textbf{Output writing scheduling.}  To correctly route the output data, the scheduling distance between the output writing and its producer must be 1. Assume that the last  addition $v_a$ in a kernel  is scheduled at $t_2$. An output writing $w$ would be scheduled at $t_3$ = $t_2$ + 1 if the modulo output buses T$_O$ at  the modulo time $t_3$ \% II are available; Otherwise, a COP $v_c$ is required for $v_a$. The $v_c$, as the  new producer of $w$, would be attempted to schedule  at every time slot after $t_3$ until the scheduling constraint  for the output writing above is met.

\subsection{Binding with Conflict graph}
Given the scheduled s-DFG and TEC,  a conflict graph is constructed to represent the binding problem, and the legal operation binding, and I/O data and internal data routing are obtained by solving the MIS on the graph. The  conflict graph construction contains:

\ding{182} \textbf{Vertices generation.}
(1) I/O reading/writing binding: $\forall$  $r$ $\in$ $V_R$ ($w$ $\in$ $V_W$)  modulo scheduled at  $m$, all the input (output) buses on $m$-th time layer of TEC are feasible. Hence, the tuples in $V_{CG}$ can be represented as  \{$(r^m$, $ibus_i^m)$ $|$ $\forall$ $r$ $\in$ $V_R$, $i$ = 1, 2, ..., $M$\} $\cup$ \{$(w^m, obus_j^m)$ $|$ $\forall$ $w$ $\in$ $V_W$, $j = 1, 2, ..., N$\}. 
(2) Operation binding: The operation binding can be expressed as quadruple $(pe^{m}_{i,j}, op^m, bus^m_x, bus^m_y)$ which can be referred to BusMap.

\ding{183} \textbf{Edges Creation.} $\forall$  $v_1$, $v_2$ $\in$ $V_{CG}$, an undirected edge is created between $v_1$ and $v_2$ if resource conflicts occur.  
Assume that $(z, xbus)$ represents the binding of  input reading ($z$ = $r$, $xbus$ = $ibus$) or output writing  ($z$ = $w$, $xbus$ = $obus$). $\forall$ $v_1$, $v_2$ and $v_3$ $\in$ $V_{CG}$, assume that $v_1$ = ($z_1^{m_1}$, $xbus_p^{m_1}$), $v_2$ = ($z_2^{m_2}$, $xbus_q^{m_2}$) and $v_3$ = $(pe^{m_3}_{i,j}$, $op^{m_3}$, $bus^{m_3}_{x}$, $bus^{m_3}_{y})$. We apply the rules proposed in BusMap for the edge creations between any two operation bindings (quadruples). The rules  related to I/O reading/writing bindings (tuples) are listed as follows:

\begin{itemize}
	\item [R1.] For $v_1$ and $v_2$, (1) $z_1$ = $z_2$; (2) $m_1$ = $m_2$, $xbus_p$ = $xbus_q$. If any of (1) or (2) satisfies, the conflict occurs.
	\item [R2.] For $v_1$ and $v_3$,  (1) ($z_1$, $op$) $\in$ $E_R$, $p$ $\ne$ $j$ or  ($op$, $z_1$) $\in$ $E_W$, $q$ $\ne$ $i$; (2) $m_1$ = $m_3$, $z_1$ $\in$ $V_R$, $p$ = $j$, $bus_y^{m_3}$ $\neq$ $\infty$ or $z_1$ $\in$ $V_W$, $q$ = $i$, $bus_x^{m_3}$ $\neq$ $\infty$. If any of (1) or (2) satisfies, the conflict occurs.
\end{itemize}

R1  respects the exclusiveness: any I/O reading/writing allocated to more than one I/O buses or any I/O bus occupied by more than one I/O readings/writings at the same time is forbidden. R2 ensures the correction of  I/O data routing, i.e., the I/O reading/writing must be bus-connected to the PEs consuming/producing the I/O data, and these buses are not allowed for bus routing.

We apply  SBTS \cite{jin_general_2015} to solve the MIS on $CG$ to get the most bindings without  conflict. If $|MIS|$ = $|V_D|$, a valid mapping is found. Otherwise, incomplete mapping happens and is then handled by the incomplete mapping processing proposed in BusMap.

\begin{table}[t]
	\centering
	\captionsetup{font=footnotesize}\caption{The features of blocks}
{
		\begin{tabular}{|c|c|c|c|c|c|c|}
			\hline
			blocks & sparsity   & $C_nK_m$ & $|V_{OP}|$ & $|V_R|$ & $|V_W|$ & $N_{FG4}$ \\
			\hline
			block1 &    0.33    &    $C_4K_6$          & 26         &   4     &   6     &     3    \\
			\hline
			block2 &    0.33    &    $C_4K_6$          & 26         &   4     &   6     &     2    \\
			\hline
			block3 &    0.42    &    $C_6K_6$          & 36         &   6     &   6     &     3    \\
			\hline
			block4 &    0.21    &    $C_4K_6$          & 32         &   4     &   6     &     3    \\
			\hline
			block5 &    0.48    &    $C_8K_8$          & 58         &   8     &   8     &     3    \\
			\hline			
			block6 &    0.62    &    $C_8K_8$          & 40         &   8     &   8     &     2    \\
			\hline			
			block7 &    0.48    &    $C_8K_8$          & 58         &   8     &   8     &     4    \\
			\hline			
		\end{tabular}
	}

\end{table}

\section{Evaluation}
\subsection{Experiment Setup}
We randomly generate 5 sparse blocks  wherein each weight in the blocks is set as zero value with a probability of 0.4.  Additionally, we choose 2 sparse blocks from sparse models of VGGNet \cite{VGG} and AlexNet\cite{AlexNet}, denoted as ``block6" and ``block7". Table 2 lists the features of the blocks. $C_nK_m$  denotes the sparse block computes $n$ channels from $m$ kernels. $N_{FG4}$ is the amount of input readings with fanout size  greater than 4.

We evaluate SparseMap against BusMap \cite{BusMap} and Zhao et al. \cite{zhao_towards_2020}    on a streaming CGRA configured as 4$\times$4 PEA, LRF with capacity of 8 and GRF with capacity of 8.

\subsection{Mapping Result Comparison}
To clearly reveal the impacts of COPs and MCIDs on the throughput of the streaming CGRA, in Table 3, we list the initiation intervals $II_0$, the number of COPs $|C|$, and the number of   MCIDs $|M|$ of the first mapping attempt as well as the finally achieved initiation intervals $II$ and speedups $S$ for all the blocks. Since  the works \cite{BusMap}  \cite{zhao_towards_2020} adopted the same scheduling heuristic\cite{llosa2001lifetime}, both works achieve the same mapping results.  
Due to no awareness of the irregular input data demands in heuristic \cite{llosa2001lifetime},  excessive COPs and MCIDs occur for \cite{BusMap}  \cite{zhao_towards_2020}, leading to  no success in the first mapping attempt for ``block1", ``block5" and ``block7". They then try to map these blocks under increased II, however, the high routing pressures caused by the  tremendous MCIDs  hinder the successful mappings for ``block5" and ``block7".  With efficient I/O data management  as well as the reconstruction of internal dependencies within adder trees,  SparseMap  not only reduces 92.5\% COPs and 46.0\% MCIDs, but also achieves the MIIs for all the blocks in the first mapping attempts. Additionally, compared with accelerating corresponding dense blocks, SparseMap  realizes the speedups  ranging from 1.5 to 2.67.

 \begin{table}[t]
	\centering
	\vspace{-0.2cm}
	\captionsetup{font=footnotesize}\caption{Mapping result comparison}
	\resizebox{1.0\columnwidth}{!}{
		\begin{tabular}{|c|c|c|c|c|c|c|c|c|c|c|c|c|c|}
			\hline
			\multicolumn{2}{|c|}{Tech.} & \multicolumn{6}{c|}{Baselines\cite{BusMap}\cite{zhao_towards_2020}} & \multicolumn{6}{c|}{SparseMap} \\
			\hline
			\multirow{2}* {blocks} & \multirow{2}*{MII} & \multicolumn{4}{c|}{First Mapping Attempt} & \multicolumn{2}{c|}{Final } & \multicolumn{4}{c|}{First Mapping Attempt} & \multicolumn{2}{c|}{Final} \\
			\cline{3-14}
			&                    &  $II_0$    & $|C|$ & $|M|$ & Success? & $II$ & $S$   &  $II_0$    & $|C|$ & $|M|$ & Success? & $II$ & $S$  \\
			\hline  
			{block1}                 &         2          &  2         &  3      & 5        &   N     & 3  &     1      &   2    & \cellcolor{cyan}{0}   & \cellcolor{cyan}{2}    & \cellcolor{cyan}{Y} & \cellcolor{cyan}{2}  &    \cellcolor{cyan}{1.5}     \\
			\hline
			block2                 &         2          &  2         &  4      & 2        &   Y     & 2  &     1.5      &   2    &   \cellcolor{cyan}{0}   & \cellcolor{cyan}{1}     &Y & 2  &      1.5    \\
			\hline
			block3                 &         3          &  3         &  5      & 11       &   Y     & 3  &    1.67       &   3    & \cellcolor{cyan}{1}   & \cellcolor{cyan}{4}   & Y & 3  &     1.67     \\
			\hline
			{block4}                 &         2          &  3         &  4      & 5        &   Y     & 3  &      1     &   2    & \cellcolor{cyan}{0}   & \cellcolor{cyan}{3}    & Y & \cellcolor{cyan}{2}  &    \cellcolor{cyan}{1.5}     \\
			\hline
			{block5}                 &         4          &  5         &  9      & 16       &   N     & \multicolumn{2}{c|}{{Failed}}           &   4    & \cellcolor{cyan}{1}   & \cellcolor{cyan}{8}   &\cellcolor{cyan}{Y} & \cellcolor{cyan}{4}  &   \cellcolor{cyan}{2}      \\
			\hline
			block6                 &         3          &  3         &  7      & 8        &   Y     & 3  &    2.67       &   3    & \cellcolor{cyan}{0}   & \cellcolor{cyan}{5}   & Y & 3  &     2.67     \\
			\hline
			{block7}                 &         4          &  5         &  8      & 16       &   N    & \multicolumn{2}{c|}{{Failed}}       & 4   & \cellcolor{cyan}{2}     & \cellcolor{cyan}{11}  & \cellcolor{cyan}{Y} & \cellcolor{cyan}{4} &    \cellcolor{cyan}{2}     \\
			\hline
			\multirow{2}* {Total} &  \multicolumn{2}{c|}{\multirow{2}*{-}} & \multirow{2}*{40} & \multirow{2}*{63} &  \multicolumn{4}{c|}{\multirow{2}*{-}} & \cellcolor{cyan}{3} & \cellcolor{cyan}{34} & \multicolumn{3}{c|}{\multirow{2}*{-}} \\
			&  \multicolumn{2}{c|}{ }                &    &    &  \multicolumn{4}{c|}{}                 & \cellcolor{cyan}{$\downarrow$ 92.5\%}           &   \cellcolor{cyan}{$\downarrow$ 46.0\%}          & \multicolumn{3}{c|}{ }\\
			\hline
		\end{tabular}
	}
	\vspace{-0.2cm}
\end{table}

\begin{table}[t]
	\centering
	\captionsetup{font=footnotesize}\caption{The impacts of  different combinations  on COPs and MCIDs}
	\resizebox{1.0\columnwidth}{!}{
		\begin{tabular}{|c|c|c|c|c|c|c|c|c|c|c|c|c|}
			\hline
			\multirow{2}* {Combinations} & \multicolumn{4}{c}{\multirow{2}* {{AIBA}}} & \multicolumn{4}{|c}{\multirow{2}* {{AIBA + Mul-CI}}} & \multicolumn{4}{|c|}{\multirow{1}* {{AIBA + Mul-CI + RID-AT}}}\\
			& \multicolumn{4}{c}{}			              & \multicolumn{4}{|c}{}                                     & \multicolumn{4}{|c|}{(SparseMap)}\\
			\hline
			blocks & $II_0$  &  $|C|$ & $|M|$ & $II$  & $II_0$  &  $|C|$ & $|M|$ & $II$ & $II_0$  &  $|C|$ & $|M|$ & $II$\\
			\hline
			block1 &  2 & 3 & 10 & 3 & 2 & 0 & 3 & 2 & 2 & 0 & \cellcolor{cyan}{2} & 2\\
			\hline
			block2 &  2 & 2 & 12 & 3 & 2 & 0 & 3 & 2 & 2 & 0 & \cellcolor{cyan}{1} & 2\\
			\hline
			block3 &  3 & 3 & 11 & 3 & 3 & 0 & 7 & 3 & 3 & \cellcolor{pink}{1} & \cellcolor{cyan}{4} & 3 \\
			\hline
			block4 &  3 & 3 & 11 & 3 & 3 & 0 & 5 & 3 & 2 & 0 & \cellcolor{cyan}{3} & \cellcolor{cyan}{2}\\
			\hline
			block5 &  5 & 3 & 23 & Failed & 4 & 0 & 13 & 4 & 4 & \cellcolor{pink}{1} & \cellcolor{cyan}{8} & 4\\
			\hline
			block6 &  3 & 2 & 14 & 3 & 3 & 0 & 11 & 3 & 3 & 0 & \cellcolor{cyan}{5} & 3\\
			\hline
			block7 &  5 & 5 & 25 & Failed & 4 & 2 & 24 & Failed & 4 & 2 & \cellcolor{cyan}{11} & \cellcolor{cyan}{4}\\
			\hline  
		\end{tabular}
	}
\end{table}

\subsection{Ablation Study}
To figure out the impacts of AIBA, Mul-CI and RID-AT techniques  on the COPs and MCIDs, we conduct an experiment in which these techniques are gradually introduced into SparseMap. From Table 4, we observe that the Mul-CI technique plays an important role to reduce the COPs and MCIDs. On the one hand, for the input readings with highly fan-out multiplications, the Mul-CI technique allocates more than one input buses for the input readings, making more multiplications obtain the input data from the input buses directly.  The Mul-CI enables more multiplications to be scheduled at the same time  as the input readings, thus avoiding the COPs.
On the other hand, the  more multiplications scheduled together  guarantee the effectiveness of the AIBA technique,  thereby  reducing the MCIDs.  After introducing the RID-AT technique, the MCIDs are further reduced for all the blocks, but with slightly increased COPs for ``block3" and ``block5". 

\section{Conclusion}
To address the issue of excessive COPs and MCIDs resulting from the  irregular input data demands inside sparse CNNs, we propose SparseMap, a mapping algorithm for sparse CNNs on streaming CGRAs. SparseMap employs an efficient I/O data management and operation scheduling \& binding to ensure optimal throughput of streaming CGRAs.  Compared to previous works, SparseMap  achieves fewer COPs and MCIDs while having the same or even better throughput.

\begin{acks}
The authors would like to thank Information Science Laboratory Center of USTC for the hardware \& software services.
\end{acks}

\bibliographystyle{unsrt}


\end{document}